\documentclass[12pt,preprint]{aastex}
\usepackage{color}

\shorttitle{Measuring the gap in HD\,169142}
\shortauthors{Honda et al.}

\begin{document}
\title{Mid-IR imaging of the transitional disk of HD169142: Measuring the size of the gap}


\author{M. Honda\altaffilmark{2}, Koen Maaskant\altaffilmark{3,4}
Y. K. Okamoto\altaffilmark{5}, H. Kataza\altaffilmark{6},
M. Fukagawa\altaffilmark{7}, 
L. B. F. M. Waters\altaffilmark{8,3}, 
C. Dominik\altaffilmark{3,10}, 
A. G. G. M. Tielens\altaffilmark{4}, 
G. D. Mulders\altaffilmark{3,9}, 
M. Min\altaffilmark{11}, 
T. Yamashita\altaffilmark{12}, 
T. Fujiyoshi\altaffilmark{13}, T. Miyata\altaffilmark{14},
S. Sako\altaffilmark{14}, I. Sakon\altaffilmark{15},
H. Fujiwara\altaffilmark{13} and T. Onaka\altaffilmark{15} 
}


\altaffiltext{1}{Based on data collected at Subaru Telescope, which is operated
by the National Astronomical Observatory of Japan.}
\altaffiltext{2}{Department of Mathematics and Physics, Faculty of Science, 
Kanagawa University, 2946 Tsuchiya, Hiratsuka, Kanagawa, 259-1293, Japan}
\altaffiltext{3}{Astronomical Institute Anton Pannekoek, University of Amsterdam, P.O. Box 94249, 1090 GE Amsterdam, The Netherlands}
\altaffiltext{4}{ Leiden Observatory, Leiden University, P.O. Box 9513, 2300 RA Leiden}
\altaffiltext{5}{Institute of Astrophysics and Planetary Sciences,
Faculty of Science, Ibaraki University, 2-1-1 Bunkyo, Mito, Ibaraki 310-8512, Japan}
\altaffiltext{6}{Department of Infrared Astrophysics, Institute of Space
and Astronautical Science, Japan Aerospace Exploration Agency,
3-1-1 Yoshinodai, Sagamihara, Kanagawa, 229-8510, Japan}
\altaffiltext{7}{Department of Earth and Space Science, Graduate School of Science, Osaka University, 1-1 Machikaneyama, Toyonaka, Osaka 560-0043, Japan}
\altaffiltext{8}{SRON Netherlands Institute for Space Research, Sorbonnelaan 2, 3584 CA Utrecht, The Netherlands}
\altaffiltext{9}{SRON Netherlands Institute for Space Research, PO Box 800, 9700 AV, Groningen, The Netherlands}
\altaffiltext{10}{Department of Astrophysics/IMAPP, Radboud University Nijmegen, PO Box 9010 6500 GL Nijmegen, The Netherlands}
\altaffiltext{11}{Astronomical Institute Utrecht, Utrecht University, PO Box 80000, 3508 TA Utrecht, The Netherlands}

\altaffiltext{12}{National Astronomical Observatory of Japan, 2-21-1
Osawa, Mitaka, Tokyo 181-8588, Japan}
\altaffiltext{13}{Subaru Telescope, National Astronomical Observatory of
Japan, 650 North A'ohoku Place, Hilo, Hawaii 96720, U.S.A.}
\altaffiltext{14}{Institute of Astronomy, School of
Science, University of Tokyo, 2-21-1 Osawa, Mitaka, Tokyo 181-0015, Japan}
\altaffiltext{15}{Department of Astronomy, Graduate School of Science, University of Tokyo, Bunkyo-ku, Tokyo 113-0033, Japan}


\begin{abstract}
\noindent
The disk around the Herbig Ae star HD\,169142 was imaged and resolved
at 18.8 and 24.5\,$\mu$m using Subaru/COMICS.  We interpret the
observations using a 2D radiative transfer model and find evidence for
the presence of a large gap.  The MIR images trace dust that
emits at the onset of the strong rise in the spectral energy distribution (SED) at 20\,$\mu$m,
therefore are very sensitive to the location and characteristics of the
inner wall of the outer disk and its dust. We determine the location of the wall
to be 23$^{+3}_{-5}$\,AU from the star.  An extra component
of hot dust must exist close to the star. We find that a hydrostatic
optically thick inner disk does not produce enough flux in the NIR and an optically
thin geometrically thick component is our solution to fit the SED.  
Considering the recent findings
of gaps and holes in a number of Herbig Ae/Be group I disks, we
suggest that such disk structures may be common in group I sources.
Classification as group I should be considered a support for
classification as a transitional disk, though improved imaging surveys are needed to support this speculation.

\end{abstract}


\section{Introduction}
Transitional disks are a class of protoplanetary disks whose
inner regions are devoid of small dust grains.  These disks have
been attracting attention recently, since such cleared inner regions
or gaps may be related to on-going planet formation.  The presence of
an inner hole/gap has been suggested for the disk around the Herbig Ae
star HD\,169142 by \cite{grady07} and \cite{meeus10}, mostly based on
the analysis of the spectral energy distribution (SED).  Both groups
conclude that the steep rise in flux at $\sim$20\,$\mu$m in the SED
reflects emission from the wall of the outer disk.  The location of
the wall (i.e. the inner edge of the outer disk) has not been determined conclusively --
\citet{meeus10} used 20 AU while \citet{grady07} suggested 44
AU.  Observations of HD\,169142 at various wavelengths have so
far not been able to constrain the inner cavity radius directly and
accurately. 

One of the limitation of SED modeling is the inability to locate and prove the existence of gaps in proto-planetary disks. For instance, gaps may not be revealed in the SED because the inner disk can partially obscure large parts of the outer disk \citep{acke09} and gaps hidden under that shadow do not leave a fingerprint in the SED. Clever SED modeling can reproduce SEDs of truly transitional disks without the need of a gap. The transitional disk LkCa 15 has an outer disk starting at 46 AU as seen at 1.4 and 2.8 mm wavelengths \citep{pietu06}. However, \cite{isella09} show that a model approach with a smooth distribution of material from a few stellar radii to about 240 AU can sometimes reproduce both the observed spectral energy distribution and the spatially resolved continuum emission at millimeter wavelengths for this object. Only from direct imaging it is possible to directly constrain the radial density structure of dust in a proto-planetary disk. 

Dust in proto-planetary disks covers a temperature range from $\sim$1500 K at the inner dust sublimation radius to a few Kelvin in the outer parts of the disk. Direct mid-infrared (MIR) imaging at 18.8 and 24.5\,$\mu$m is most sensitive for 100-150\,K blackbody dust, though strong contributions from dust elsewhere in the disk at lower and higher temperatures can also be present. If the disk has a very strong NIR component such as HD\,135344 \citep{meeus01}, the inner region may still be strongly represented in the center of the image at 18 micron. On the other hand, for disks with a larger gap or an inner hole, the image size will be a sensitive tracer to the location of the inner radius of the outer disk \citep{verhoeff11}.   

In this paper, we present direct MIR imaging
observations of HD\,169142 at 18.8 and 24.5\,$\mu$m, using COMICS on the
8.2m Subaru telescope.  We confirm the result found by \cite{marinas11} and find that the
 disk shows extended emission at 18\,$\mu$m. In addition we find that the disk  
 is also resolved at 24.5\,$\mu$m.  By constructing a radiative transfer disk model that fits both the SED and
the imaging results, we find that the size of the source at MIR
wavelengths is most naturally explained by a disk with a large inner gap,
i.e. transitional disk.  
We discuss the structure of the disk, and the implications for
its nature.

\section{Observations and Data Reduction}
Observations were conducted using COMICS \citep[Cooled Mid-Infrared
Camera and Spectrometer;][]{kataza00,okamoto03,sako03} on the 8.2 m
Subaru Telescope on Mauna Kea, Hawaii.  HD\,169142 was observed using
the Q24.5-NEW ($\lambda$=24.5\,$\mu$m,
$\Delta\lambda$=0.75\,$\mu$m) and
Q18.8($\lambda$=18.8\,$\mu$m,$\Delta\lambda$=0.9\,$\mu$m)
filters.  The plate scale of the COMICS camera is 0.13 arcsec per
pixel. The chopping throw was 10 arcsec and the position angle of
chopping direction was 0 degree. The chopping frequency was 0.45 Hz.
The total integration time for Q24.5 and Q18.8 observations were 802\,s
and 360\,s, respectively.
Just before and after observing HD\,169142, we took data of PSF
reference and photometric standard stars.  We used $\delta$ Oph for
Q24.5 and $\alpha$\,Her for Q18.8.  The total integration times for
the reference stars were 243\,s and 83\,s, respectively.  A summary of
the observations is given in Table 1.

We processed the data using a shift-and-add method to improve the
blurring caused by atmosphere, tracking errors, misregistration, etc. 
The imaging data consists of 0.98\,s on-source
integration frames.
First, the thermal background and the dark current signals were
removed through the subtraction of the chopped pair frames. The object and the PSF star
are bright enough to be recognized even in 0.98\,s chop-subtracted
frames, so we searched the centroid of the object. 
We then shifted the frames so as to align the centroid position, 
and summed up the frames.  However, we excluded frames which
were blurred by atmospheric seeing. We rejected frames whose radius of 78.4\% encircled 
energy ($r_{78.4}$) is larger than thrshould radius ($r_c$) which covers 95\% of the wavefronts whose
Strehl ratio is 0.9.
These thrshould values are determined as $r_c = 1.131''$ at 18.8\,$\mu$m and $r_c = 1.482''$ at 24.5\,$\mu$m
by Monte Carlo simulation of ideal unresolved point source.
Due to this rejection of lower quality
data, the effective integration time of HD\,169142 was reduced to 345\,s
and 215\,s for Q24.5 and Q18.8, respectively.  The same procedure was
applied to the PSF images taken before and after HD\,169142, where the
effective integration time became to 195\,s and 83\,s,
respectively. Since HD\,169142 shows marginal extention compared to the PSF and 
the rejection criterion $r_c$ is determined for the ideal unresolved point source, 
frame rejection rates of HD\,169142 are higher than those of PSF stars.
It means that the effective rejection criterion is slightly more stringent for HD\,169142 
than for PSF stars, however, it does not overestimate the extention of HD\,169142.

\begin{figure}
\epsscale{1.2}
\plottwo{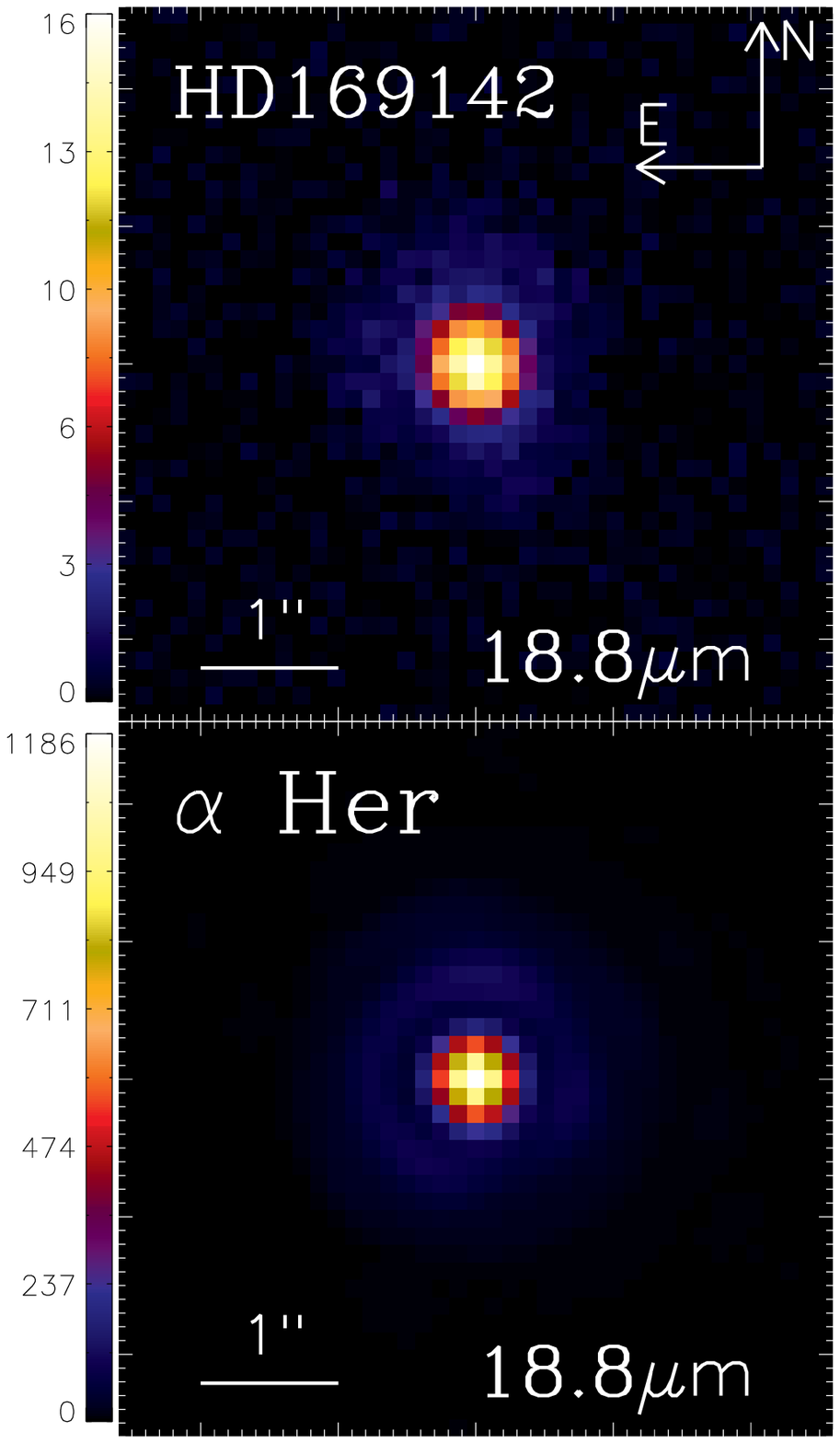}{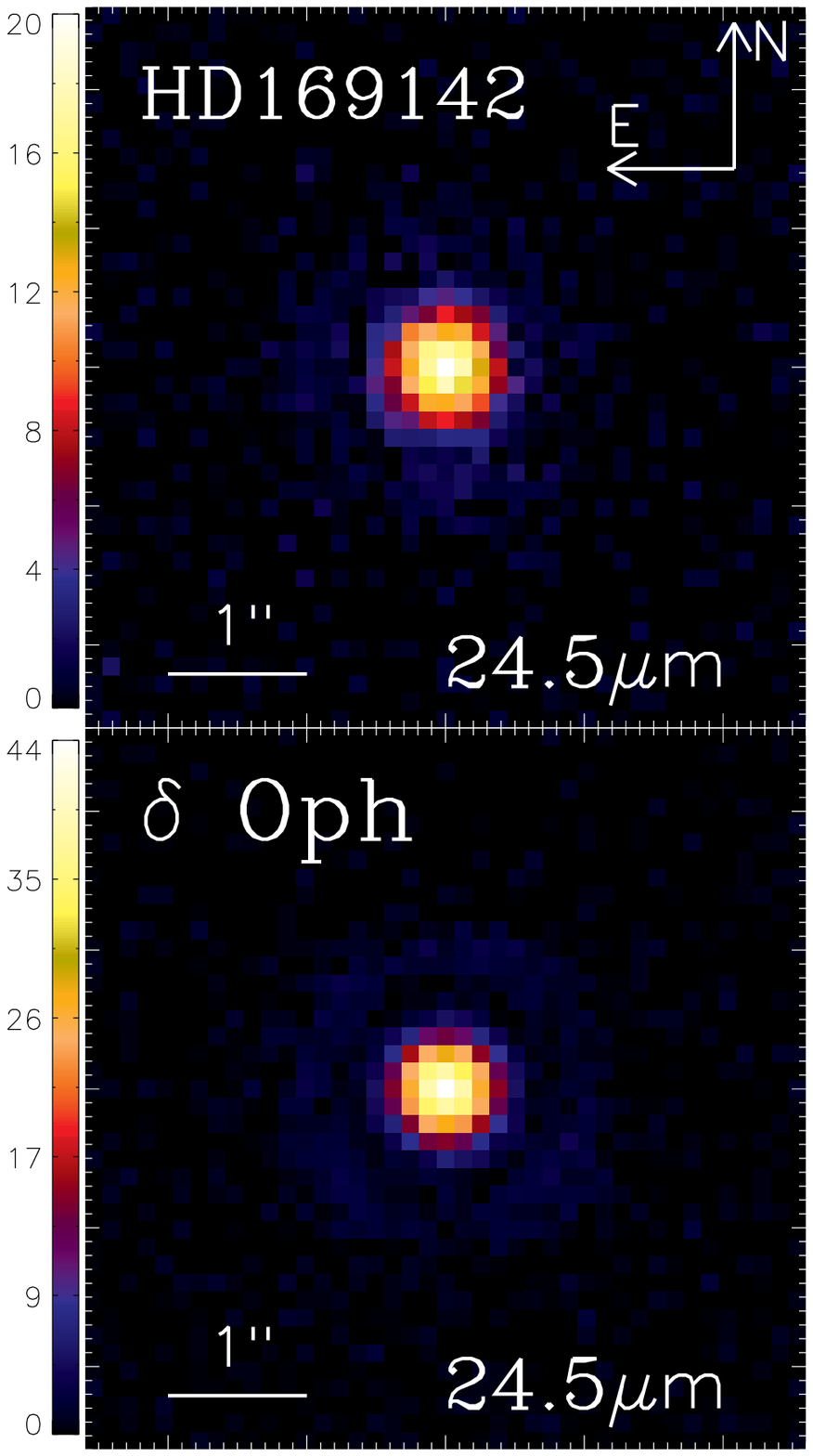}
\caption{Shift-and-added image of HD 169142 (top) and PSF star (bottom) at 18.8 and 24.5 \,$\mu$m. 
Brightness unit is in Jy/arcsec$^2$ and scaled from 0 to the peak
 value in the image. North is up and the east is
 to the left \label{obsimages}}
\end{figure}

For flux calibration we used template spectrum provided
by \cite{engelke06}. Using the standard stars observed over the two nights, 
an airmass correction was applied for Q24.5 photometry. For
Q18.8 photometry, we could not find a significant airmass dependence during the
night.  A standard aperture photometry was applied and the resultant
flux density of HD\,169142 was $10.5 \pm 0.4$\,Jy at 18.8\,$\mu$m and
$13.0 \pm 0.5$\,Jy at 24.5\,$\mu$m. 
The final images of the PSF and HD\,169142 are shown in Figure \ref{obsimages}.

\section{Results}
\subsection{Source size of HD\,169142 in 18.8 and 24.5\,$\mu$m}
The azimuthally-averaged radial brightness profiles of HD\,169142 and
the PSF stars at 18.8 and 24.5\,$\mu$m are shown in Figure \ref{fig:RBP}.
It is clear that HD\,169142 is extended at these wavelengths. At
18.8\,$\mu$m the direct FWHM of the shift-and-added images of
HD\,169142 and of the PSF stars were 0.604''$\pm$0.017'' and 0.493''$\pm$0.006'', respectively.
At 24.5\,$\mu$m, the corresponding numbers are 0.680''$\pm$0.034'' and 0.628''$\pm$0.007'' (see
Table 1). It is surprising that the size of HD169142 is not 
increasing with wavelength. For a continuous flaring disk, one would 
expect the size of the image to scale roughly with the PSF \citep{meijer07}.
Our images therefore give a first indication that the radial density structure is not continuous.
The FWHMs of the PSF references are comparable to the
predicted value of the diffraction limited performance of the
telescope.  As a rough estimate of the intrinsic source size of
HD\,169142, we applied the quadratic subtraction method as described by
\cite{marinas11}.  The derived source FWHMs were 0.349''$\pm$0.014'' at
18.8\,$\mu$m and 0.261''$\pm$0.025'' at 24.5\,$\mu$m.  Since \cite{marinas11}
observed the FWHM of this source to be 0.32''$\pm$0.05'' at
18.0\,$\mu$m, our measurement at 18.8\,$\mu$m is in good agreement with their results within the uncertainties.

\section{Modeling}
\subsection{Observational constraints}

Complementary views of the disk structure and the dust properties of
HD\,169142 are obtained by adopting photometric data at various wavelengths from the literature
\citep{vandenAncker97,cutri03,zacharias04,sylvester96,meeus10} and MIR
spectra from ISO \citep{meeus01} and Spitzer/IRS
\citep{juhasz10}. These datasets reflect the disk brightness as a
function of wavelength and are a result of the disk structure defined
by the density distribution, composition, inclination, and inner and
outer radius of the disk and by the properties of the central star. An
outer disk radius of 235 AU is taken from \citet{panic08}.  An inner
disk radius of 0.1 AU is set at the location where the dust reaches
1500 K and starts to sublimate.  With an inclination of 13$^{\circ}$
\citep{raman06}, the system is nearly pole-on and, therefore, 
deviations from axial symmetry caused by projection effects can be neglected in our images.  The stellar
spectrum is described by a Kurucz model with an effective temperature
of 8200 K \citep{dunkin97}, luminosity of 15.3 L$_\odot$ and extinction
A$_{V}$=0.46 \citep{vandenAncker99}.  The system is set at a distance
of 145 pc \citep{dezeeuw99}.  All stellar and disk parameters used in
this paper are shown in table \ref{tab:parameters}.

\subsection{Radiative transfer code MCMax}

We construct a geometrical model of the disk of HD\,169142 and fit it to both
the radial surface brightness profile (RBP) of our observed images and to the SED.
For the model we use the radiative transfer code MCMax \citep{min09},
which is able to solve temperature and density structures in very
optically thick circumstellar disks.  This code has been compared to
other radiative transfer codes by \cite{pinte09} and has been applied
successfully in previous studies \citep[e.g.][]{verhoeff11,mulders11}.
We assume an axisymmetric dust distribution in which the optical
properties of the grains are computed using a distribution of hollow
spheres \citep[DHS;][]{min05}.
We consider all dust species in the disk to be in thermal contact, and
calculate their respective opacities from the optical constants as if
they were separate particles.  Furthermore we assume the gas
temperature to be set by the dust temperature. We adopted grains with sizes
according to the power-law $f(a) \propto a^{-3.5}$.  The power-law
index is that of interstellar grains
\citep{Mathis77} and is consistent with collisional fragmentation considerations
\citep{Hellyer70}.  Under the assumption
that the system is in hydrostatic equilibrium, the vertical density distribution of the
disk is solved by iterating the density and temperature structures
until they become self-consistent \citep[e.g.][]{dullemond07}.

We assume a grain population consisting of 30\% carbon and 70\% silicates 
(we refer to \cite{mulders11} for references of the optical properties of this dust composition). 
The contribution of warm small ($<$ 10
$\mu$m) amorphous silicate grains to the spectrum must be low, since
the flux level at 10 $\mu$m is very low compared to that at 20 $\mu$m.  We
model this change in flux level by replacing the amorphous silicate dust in the inner
disk with dust with a higher continuum opacity.
We here use carbon, but other dust species such as metallic iron may
give the same result. This treatment is consistent with \citealt{VanBoekel04a} who find 
that grain growth in the innermost regions has proceeded further than in
the outer disk regions. We have not attempted to fit the PAH
features in detail, though we implement them into our model to fit the SED.
Scattered light contribution is also included and the fraction of the scattered light to the 
total flux density in 18.8 and 24.5 $\mu$m is 22--26\% in this model.

\begin{figure}[]
\includegraphics[scale=0.85]{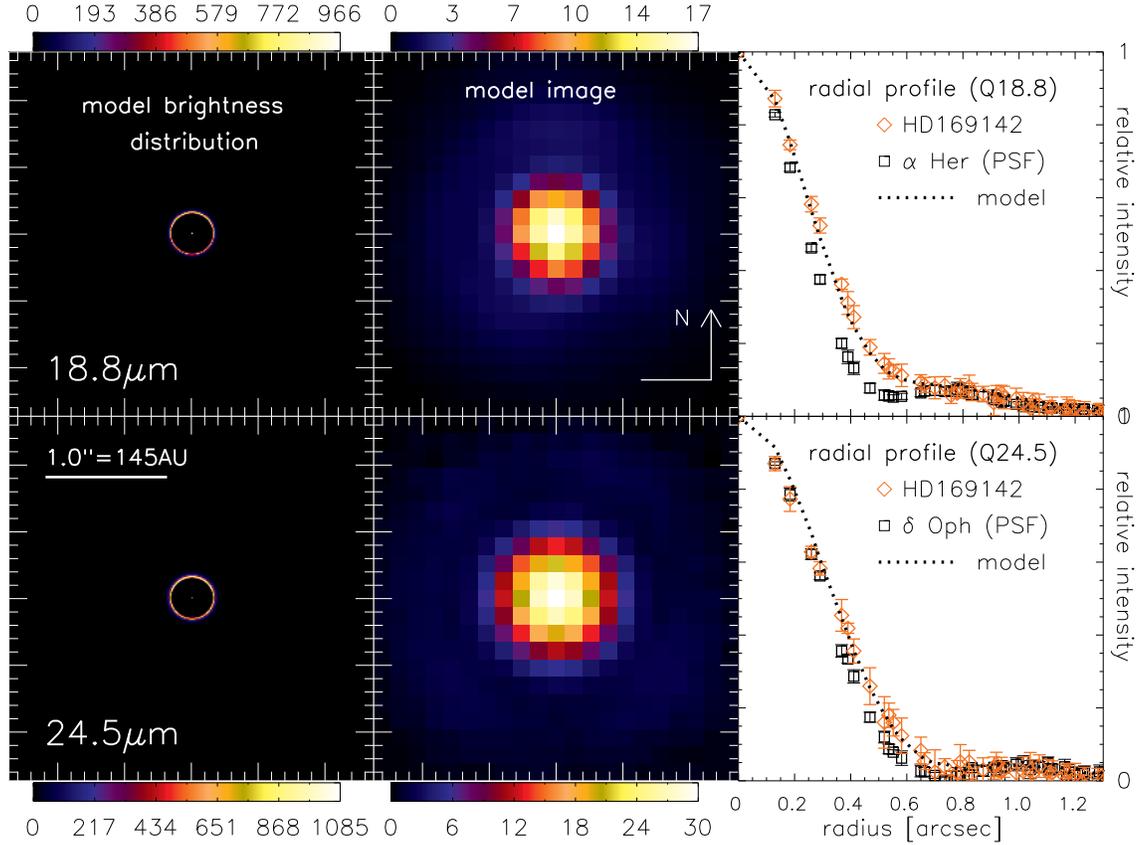}
\vspace{10 mm}
\caption{\label{fig:RBP} Left and middle panels are model brightness distribution of final model and
model image after convolved by Subaru/COMICS PSF (middle), respectively, shown in Jy/arcsec$^2$. 
Right panels represents the peak-normalized azimuthally-averaged 
radial brightness profiles relative to the centroid of the image (right).
Top panels are in the Q18.8 and the bottom panels are in the Q24.5 filters.
In the radial profile plots (right), the observation of HD\,169142 is shown by the red diamonds, 
and the corresponding PSF of the calibrators is shown by the black squares. 
The final model convolved with the PSF is shown by the black dotted line.}
\end{figure}

\begin{figure}[!h]
\includegraphics[scale= 0.8,angle=90]{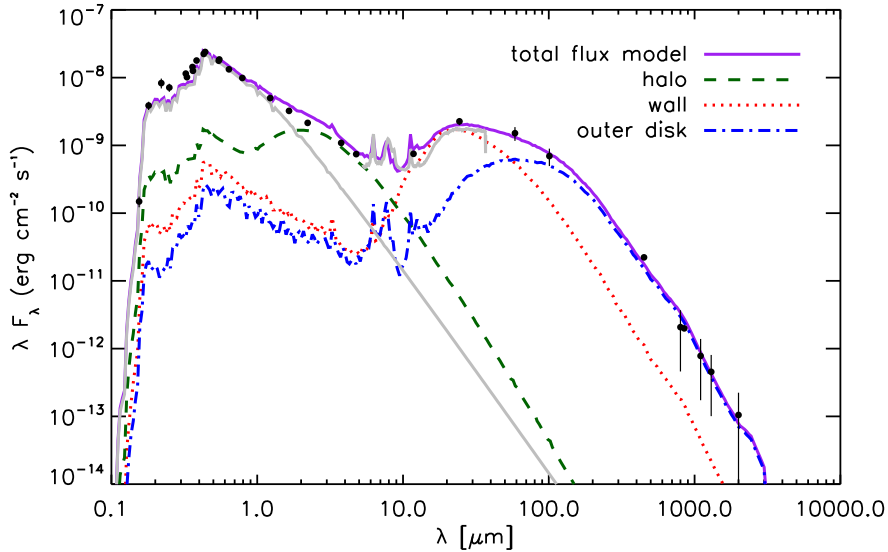}
\caption{\label{fig:SED}Spectral energy distribution of HD\,169142.  The solid grey lines
  represent the Kurucz and the Spitzer/IRS [5.3 - 37
  $\mu$m] spectra respectively. The dark green dashed line shows the geometrically thick, optically
thin halo-like dust cloud component. The red dotted line gives the flux contribution 
  from the innermost part of the outer disk (23-26 AU). The flux from the remainder of the outer disk (26-237 AU) is given by the dashed-dotted blue line. The sum of the components is given by the purple solid line. It is clear that at both 18.8 and 24.5 $\mu$m, the SED is dominated by flux coming from the inner edge of the outer disk. }
\end{figure}

\begin{figure}[!h]
\includegraphics[scale= 0.4,angle=90]{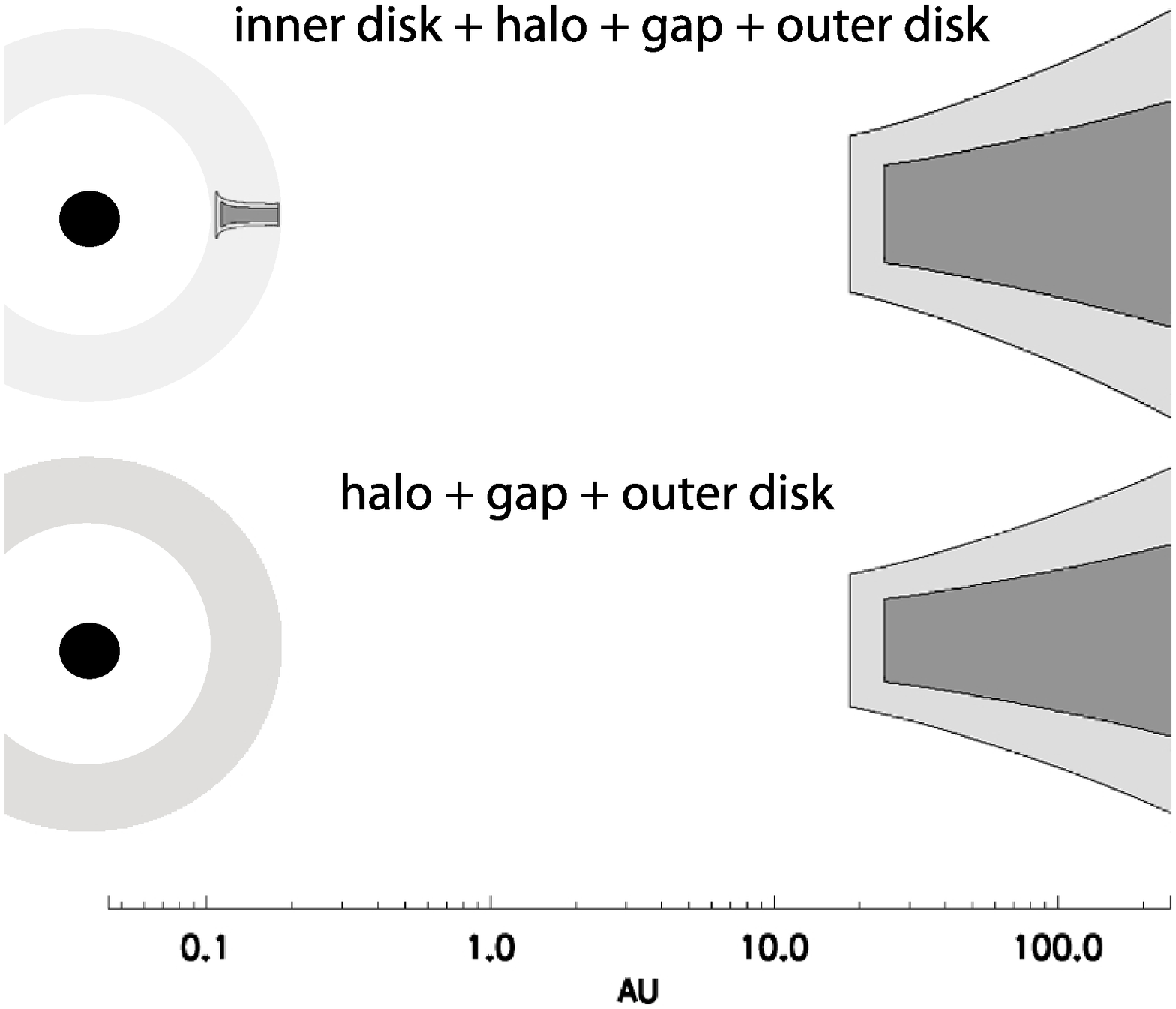} 
\caption{\label{fig:cartoon} Sketch of our final model with two possible structures for the inner regions. Top: the model with a flat inner disk, a optically thin halo and an outer disk. Bottom: the model with a denser but still optically thin halo, and an outer disk in which the scaleheight of the dust is decreased. Both geometries give an equally good fit to the SED and the images.  Although we can not distinguish between these models with our dataset, we choose the bottom model as our final model described in the text.  }
\end{figure}

\begin{figure}[!h]
\includegraphics[scale= 0.55, angle=90]{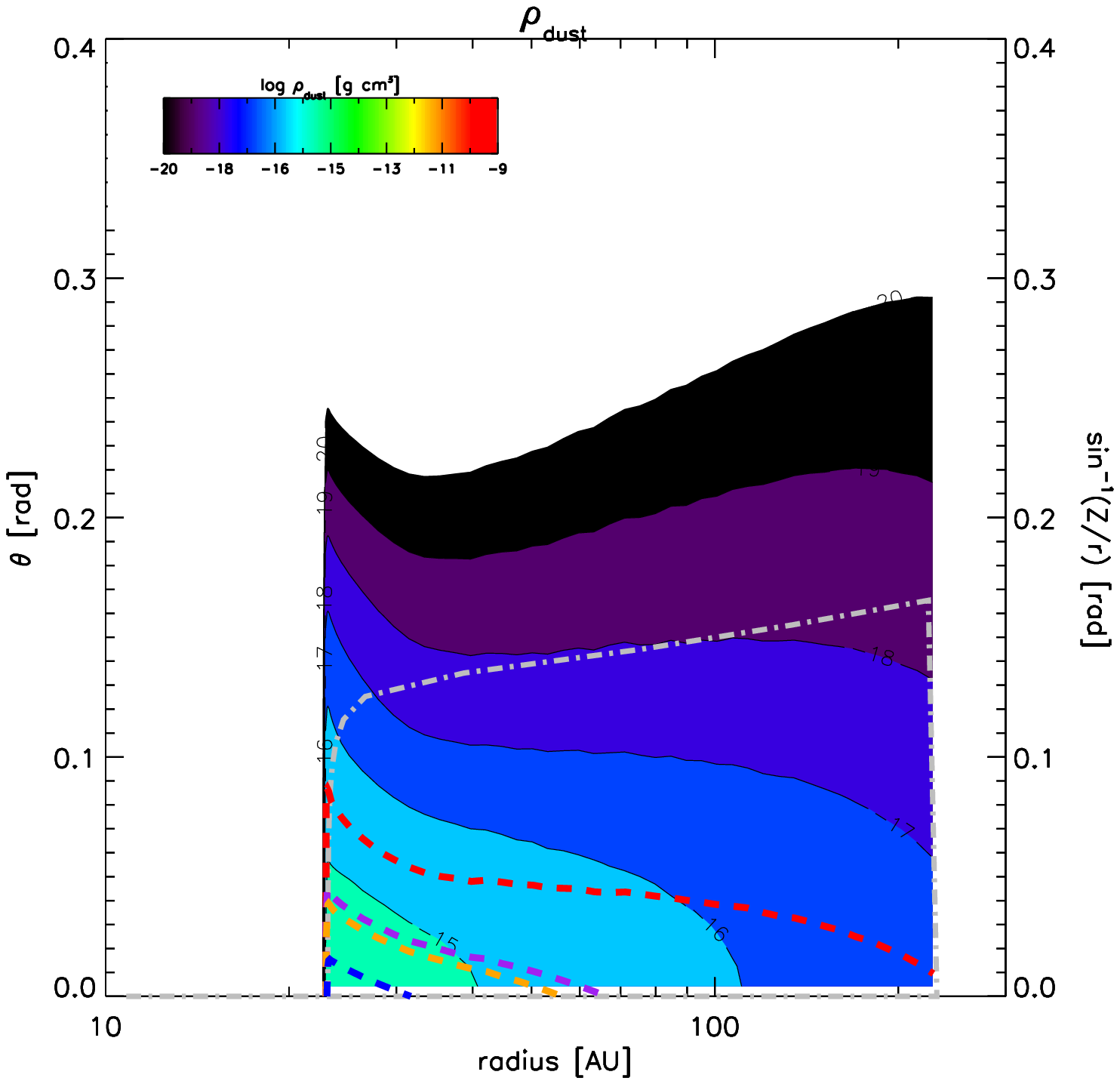} 
\includegraphics[scale= 0.55, angle=90]{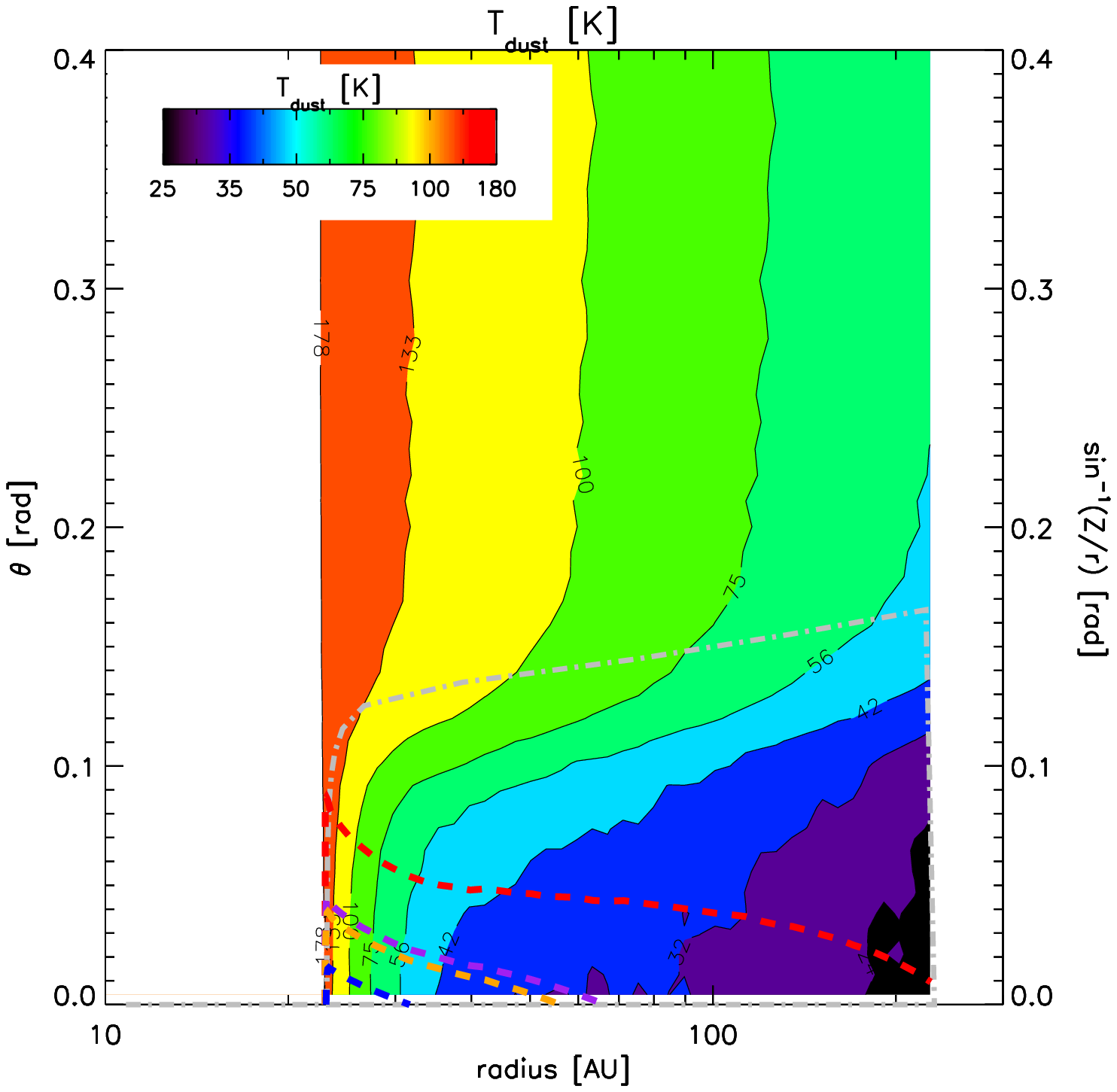} 

\caption{\label{fig:dens}Left: the dust density $\rho_{\rm dust}$ in g cm$^{-3}$ of the outer disk as a function of radius and angle $\theta$ for our best fitting model. The lower and upper density cut-offs are 10$^{-20}$ and 10$^{-9}$ g cm$^{-3}$ respectively.  Right: the temperature structure of the outer disk as a function of radius and angle $\theta$ for our best fitting model. The inner edge of the outer disk starts at a temperature of 180 K. For small $\theta$ the angle is approximately Z/r. The grey dot-dashed line shows the $\tau$=1 surface in radial direction in scattered light ($\lambda$=0.5 $\mu$m). The dashed lines represent the $\tau$=1 surfaces as seen perpendicular from the disk. From top to bottom the colors are red, purple, orange and blue and represent the wavelengths 0.5, 18.8, 24.5, 70 $\mu$m. }

\end{figure}

\subsection{The best fit model}




We fit the SED as well as the 18.8 and 24.5 $\mu$m images simultaneously by using the disk parameters shown in Table \ref{tab:parameters}.  The RBPs and SED of our final best fit model are shown by the the purple dashed lines in Figures \ref{fig:RBP} and \ref{fig:SED}. A cartoon sketch is given in Figure \ref{fig:cartoon}. In this section we will first discuss the geometrical characteristics of the location of the gap and the outer disk properties. Thereafter we discuss the constraints we find to the inner disk structure of the best fitting model.

\subsubsection{Fitting the SED}
Previous SED modeling by  \cite{meeus10} and  \cite{grady07}suggested a gap in the disk with an outer radius of 20 and 44 AU respectively. We confirm and improve on this result by our test modeling for HD169142: disks without a gap fail to fit the SED, despite various attempts within the physically reasonably parameter space. So already the SED fitting gives a strong indication for the presence of a gap (confirming the results by \citet{meeus10} and  \citet{grady07}, even though its size is not well constrained at all from the SED alone. We incorporate this result and include a
gap in our model. The effect of the discontinuity in the radial density structure is
that a large fraction of the stellar light now falls on the vertical wall at the inner edge of the outer disk. Emission from this 
wall is clearly distinguishable in the SED (Figure \ref{fig:SED}, illustrated by the 
red dotted-line). In addition, Figure \ref{fig:SED} shows that the flux contribution 
from the inner edge (between 23-26 AU) dominates at 18.8 and 24.5 $\mu$m  and that flux contributions from the inner regions and from the outer disk are 
negligible at these wavelengths. Therefore, our Subaru/COMICS images at 18.8 and 24.5 $\mu$m are sensitive tracers of the radius where the outer disk starts. Another insight to understand the dominance of the wall is given by the radial density and temperature plots of the outer disk of this model in Figure \ref{fig:dens}. The dotted lines show that the densest regions are largely optically thick. Therefore the temperature of the disk mid-plane decreases rapidly as a function of radius. Thus 18.8 and 24.5 $\mu$m photons are mostly produced at the inner few AU radii of the outer disk. 

The radius where the outer disk starts is not well constrained by SED modeling. Because the disk structure is degenerate by uncertainties of many parameters like the stellar temperature and luminosity, the interstellar extinction, the grain size population of the dust, the vertical density structure and so on.  Only by fitting our model to the resolved MIR observations we can unambiguously confirm the existence of a gap and constrain the location of the wall of the outer disk.

\subsubsection{Fitting the MIR-images, constraining the location of the wall}

After convolving our best fit model with the PSFs of the calibrators, we find that the imaging
data is fitted best with an inner edge of the outer disk of 
23$^{+3}_{-5}$\,AU (see Fig.\ref{fig:RBP}). The uncertainty on this radius is determined by fitting the uncertainty of the azimuthally averaged flux level at $\pm$1 sigma. This
result is consistent with a wall radius of 20 AU \citep{meeus10} but
excludes the model with a wall at 44 AU  \citep{grady07}.

We have performed a small parameter study where we have examined the robustness of our final model to fit the RBPs at MIR wavelengths.
We have varied the following parameters: the surface density powerlaw between \{-0.5, -1.5\}, the grain size distribution powerlaw \{-2.5, -4.5\}, the dust settling factor \{0.5, 1\}, the stellar temperature \{8000 K, 8500 K\} and luminosity \{13 L$_{\odot}$, 17 L$_{\odot}$\} and the distance \{130 pc, 160 pc\}. While the RPB only vary slightly in this parameter study (all within the 1 sigma error of the measurements), the SED fit becomes significantly worse.  We conclude that the radius of the wall is the only free parameter which significantly influences the broadness of the RBP. Therefore our fit to the RBP is robust in constraining the radius of the inner edge of the outer disk at 23$^{+3}_{-5}$ AU.

\subsubsection{The inner disk}
\label{sec:innerdisk}

In this section we explain in more detail the implications for the inner disk structure.
Modeling shows that an optically thick hydrostatic inner disk does not produce enough flux to fit the SED. Scaling up the vertical height of the inner disk would not be a solution because if the inner disk is higher than  Z/r $\sim$ 0.1 it covers the outer disk for more than $\sim$50\% in direct and scattered light. This inner disk then casts a too large shadow on the outer disk to be able to reproduce the SED at wavelengths longer than $\sim$10 $\mu$m. To solve this problem we include a geometrically thick, \emph{optically
thin} halo-like dust cloud to the inner regions ($<$ 0.5 AU, see Figure \ref{fig:dens}) to fit the
SED in the NIR and to avoid a shadowing effect on the outer disk.  The physical origin is not certain at this moment,
though other observational studies show that an optically thin
component close to the star provides a good fit as well
\citep{grady07,benisty11,verhoeff11,mulders10}. Furthermore, fitting the Herbig stars' 
median SED requires such a halo \citep{mulders12}.
Further high spatial resolution observations such as infrared interferometry will be necessary to reveal the inner disk structure. 

If there is no shadowing effect from the inner disk, the wall of the outer disk is too luminous.  To fit the SED again, we can either settle the grains in the outer disk by scaling down the dust density scaleheight by a factor of 0.6 or we can scale down the abundance of small ( $<$ 1 $\mu$m) grains by one order of magnitude. These adjustments are consistent with dust settling and grain growth
\citep{DullemondDominik05}, however, it is uncertain to what extent these effects are present in the disk. The influence of other effects like dynamical scattering of larger dust grains may also play a role in the vertical grain size distribution. Alternatively, we can \emph{combine} a halo-like dust cloud and a flat inner disk and consequently perform no settling or removal of small dust grains. So we conclude that there are two possible inner disk geometries (as illustrated in the cartoon in Figure \ref{fig:cartoon}) which give an equally good fit to the SED and the RBP.

Without an optically thick inner disk, the RBP is still reproduced with the inner radius of the outer disk at
23$^{+3}_{-5}$ AU.  The final model can be interpreted as an inner halo-like dust cloud 
\citep{krijt2011} and a gas-rich outer disk structure, so truly a
transitional disk.


\section{Discussion}

\subsection{Inner hole/gap are common for Meeus's group I sources ?}

We have shown observational evidence for the presence of an inner hole
or gap in the HD\,169142 disk.  A similar structure is also indicated
for other Herbig Ae/Be stars such as AB\,Aur \citep{honda10},
HD\,142527 \citep{fukagawa06,fujiwara06,verhoeff11}, HD\,135344
\citep{brown09}, HD\,36112 \citep{isella10}, HD\,100546
\citep{bouwman03,benisty10,mulders11}.  All of these
objects have been classified as group I according to the definition introduced by
\cite{meeus01}.  Since the strong far infrared excess is a
characteristic of group I sources, we speculate that this 
component may come from
the emission from the inner-edge of the outer disk.  Consequently, we
suggest that the inner hole or gap structure may be common
for group I sources.  If this is true, the hypothesis that the
group I flared disks evolves into the group II flat disks due to 
dust sedimentation should be reconsidered, because the difference of
group I and II is not only the degree of disk flaring or grain growth, but also the
presence of cleared inner regions (holes and gaps) in the disk.

As we have shown in the above analysis, the spatial extent of a Herbig Ae/Be disk at MIR wavelengths, often
described by a FWHM, is strongly influenced by the position
and temperature of the wall.  Previous studies showed that group I
sources are likely to be more extended than group II sources in the
MIR wavelengths, and suggested that the disk geometry (flaring or flat
disk) may play an important role in the thermal structure and MIR
emission of the disk \citep{leinert04,liu07,marinas11}. 
Our 24.5\,$\mu$m imaging survey of Herbig Ae/Be stars with
Subaru/COMICS (Honda et al. in preparation) also confirms that many group I sources are extended, 
however, their spatial extent shows a great diversity, from marginally to remarkably extended. 
Such diversity in the spatial extent in the MIR can be understood in terms of the distance, the 
inner disk structure and the location of the wall of the outer disk. 
In fact, the temperature of the wall both in the AB\,Aur and
HD\,142527 systems appears to be relatively cool (70-100 K) and its 
wall radius also tends to be at some distance away from the central star
(~100 AU in AB Aur; ~170 AU in HD\,142527 \citep{fujiwara06}).  The
wall in HD\,169142 is at a smaller distance (23AU), thus the spatial extent is 
not so large compared to that of AB Aur and HD142527.  We suggest that not only the
geometry (flared or flat) of the disk, but also the inner gap and a
wall-like inner edge of the outer disk are important for understanding
the spatial extent of Herbig Ae/Be disks in the MIR.

\subsection{Origin of the inner hole of HD\,169142 disk}

Gaps seem to be a common characteristic of the group I flaring disks
\citep{honda10,fukagawa06,fujiwara06,verhoeff11,brown09,isella10,bouwman03,benisty10,mulders11}.
An inner hole or gap in the disk has been explained by several
mechanisms, including (1) photoevaporation of the disk \citep{alexander06}, 
(2) geometric shadowing \citep{dullemond04},
(3) grain growth in the inner disk causing a lower dust opacity zone \citep{vanBoekel05},
and (4) the presence of another body in the disk that dynamically
creates a gap and decouples the inner disk from the outer disk
\citep{augereau04}. We prefer solution (4) since this also provides a
natural explanation for a inner halo-like dust cloud in the inner
disk. The presence of a halo likely requires some dynamical
interactions with planetary bodies since new generations of dust can
than be created in situ up to large disk heights \citep{krijt2011}.
Grain growth can not explain the presence of a gap and a wall.
Furthermore, our test models show that geometric shadowing
(i.e. varying scaleheights, surface density powerlaws) of a part of
the disk at several AU can not reproduce the SED and the
RBP of the images. We recognize that planetary
companions have not yet been detected in the transitional disk around HD\,169142,
though recent direct detection of the planets around A type stars
(e.g. HR8799; \cite{marois08}, $\beta$Pic \cite{lagrange10}) leads support
for such hypothesis. Thus we suggest that the HD\,169142 system is an excellent 
candidate to look for newly-formed planets in the protoplanetary
disk.

\section{Conclusions}

We arrive at the following conclusions:

\begin{enumerate}
\item\label{item:1} HD\,169142 is extended at MIR wavelengths with the Subaru/COMICS instrument.  The
  derived source sizes from quadratic PSF reference subtraction are
  0.349''$\pm$0.014'' at 18.8\,$\mu$m and 0.261''$\pm$0.025'' at 24.5\,$\mu$m, respectively.
\item\label{item:2} The observed sizes as well as the steep rise of
  the SED near 20\,$\mu$m require a disk model with an inner hole the size
  of 23$^{+3}_{-5}$\,AU, making HD\,169142 a transitional disk.
\item\label{item:3} HD\,169142 is one of a growing number of Herbig Ae
  disks that are classified as group I source and that \emph{also}
  require the presence of a large inner gap.  It appears that many if
  not all group I sources may be strong candidates for classification
  as transitional disks.
\end{enumerate}

\acknowledgments

We are grateful to all of the staff members of the Subaru Telescope.
We also thank Ms. Hitomi Kobayashi and Dr. Yuji Ikeda at Kyoto-Nijikoubou Co., Ltd..
KM is supported by a grant from the Netherlands Research School for Astronomy
(NOVA).
This research was partially supported by KAKENHI
(Grant-in-Aid for Young Scientists B: 21740141) by the Ministry of
Education, Culture, Sports, Science and Technology (MEXT) of Japan.



\begin{deluxetable}{cccccc}
\tabletypesize{\scriptsize}
\tablecaption{Observations Summary}
\tablewidth{0pt}
\tablehead{
\colhead{object} &
\colhead{filter} & \colhead{Date(UT)} & \colhead{Integ. time [s]} &
\colhead{AirMass} & \colhead{Direct} \\
\colhead{} &\colhead{} & \colhead{} & \colhead{used/total (\%)} &
\colhead{} & \colhead{FWHM} }
\startdata
$\delta$Oph & Q24.5 & 2004/07/11 &  195/243 sec (80 \%) & 1.254-1.354 & 0.628''$\pm$0.007'' \\
HD\,169142   & Q24.5 & 2004/07/11 &  345/802 sec (43 \%) & 1.541-1.542 & 0.680''$\pm$0.034'' \\
$\alpha$Her & Q18.8 & 2004/07/12 &  83/83 sec (100 \%)  & 1.081-1.115 & 0.493''$\pm$0.006'' \\
HD\,169142   & Q18.8 & 2004/07/12 &  215/360 sec (60 \%) & 1.553-1.559 & 0.604''$\pm$0.017'' \\ \tableline
\enddata
\end{deluxetable}

\begin{deluxetable}{ccc}
\tabletypesize{\scriptsize}
\tablecaption{\label{tab:parameters}Parameters of HD\,169142 system used in our best-fit model}
\tablewidth{0pt}
\tablehead{
\colhead{parameter} &
\colhead{value} & \colhead{Remarks}}
\startdata
Spectral type & A5Ve & \cite{dunkin97}\\
Extinction A$_{V}$ & 0.46$\pm$0.05 &  \cite{vandenAncker99}\\
log g & 4.22 &  \cite{vandenAncker99} \\
Temperature & 8200K & \cite{dunkin97} \\
Distance & 145$\pm$15 pc & \cite{dezeeuw99} \\
Age & 6$^{+6}_{-3}$ Myr & \cite{grady07}\\
Stellar luminosity & 15.33$\pm$2.17 L$_\odot$ & \cite{vandenAncker99} \\
Stellar Mass & 2.28$\pm$0.23 M$_\odot$ &  \cite{vandenAncker99} \\
Stellar Radius & 1.94$\pm$0.14 R$_\odot$ &  \cite{vandenAncker99} \\
Gas disk mass  & (0.16--3.0)$\times$10$^{-2}$ M$_\odot$ & \cite{panic08} \\
Dust disk mass & 4$\times$10$^{-4}$ M$_\odot$ & fit to the sub mm photometry \\
Inclination & 13$^o$ & \cite{raman06,dent05}\\
accretion rate & $\leq$10$^{-9}$M$_\odot$ yr$^{-1}$ & \cite{grady07}\\
$R_{halo}$ & 0.1-0.2 AU & geometrically heigh, optically thin component to fit the NIR \\
$R_{in}$ & 23$^{+3}_{-5}$ AU & fit to RBP of Subaru/COMICS data\\
$R_{out}$& 235 AU & \cite{panic08}\\
surface density exponent & -1.0 & Hydrostatic equilibrium \\
Particle size & a =\{0.03$\mu$m,1cm\} & powerlaw distribution of -3.5\\
Silicates & 70\% & similar to \cite{mulders11}\\
Amorphous carbon & 30\% & \cite{zubko96}\\
$M_{PAH}$ & 0.45 $\times$ 10$^{-7}$ M$_\odot$ & uniform PAH distribution \\
$M_{halo}$ & 0.28 $\times$ 10$^{-10}$ M$_\odot$ & only carbon \\
$M_{disk}$ & 0.3 $\times$ 10$^{-3}$ M$_\odot$ & mass of grains a =\{0.03$\mu$m, 1cm\} in the disk  \\

\enddata
\end{deluxetable}


\end{document}